\documentclass[manuscript,nonacm,screen]{acmart}

\newcommand{\paperTitle}{Towards a Periodic Table of Computer System Design Principles}
\newcommand{\paperKeywords}{}
\newcommand{\paperAuthors}{Joy Arulraj}

\usepackage{graphicx}
\usepackage{url}
\usepackage{xspace}
\usepackage{array}
\usepackage{multirow}
\usepackage{booktabs}
\usepackage{balance}
\usepackage{wrapfig}
\usepackage{subfig}
\usepackage{listings}
\usepackage{caption}
\usepackage[capitalize,noabbrev,nameinlink]{cleveref}
\usepackage{etoolbox}

\definecolor{websky}{HTML}{4689CC}

\hypersetup{%
    pdfauthor = {\paperAuthors},
    pdftitle = {\paperTitle},
    pdfkeywords = {\paperKeywords},
    bookmarksopen = true,
    colorlinks = true,
    citecolor = websky,
    linkcolor = websky,
    urlcolor = websky,
    pdfborder = {0 0 0}
}

\usepackage[ruled,linesnumbered,noend]{algorithm2e}

\SetCommentSty{mycommfont}
\SetAlFnt{\small}
\SetAlCapFnt{\small}
\SetAlCapNameFnt{\small}

\crefname{lstlisting}{listing}{listings}
\Crefname{lstlisting}{Listing}{Listings}
\crefname{figure}{Fig.}{Figs.}
\Crefname{figure}{Fig.}{Figs.}
\crefformat{section}{\S\xspace#2#1#3}
\crefformat{subsection}{\S\xspace#2#1#3}
\crefformat{subsubsection}{\S\xspace#2#1#3}
\crefrangeformat{section}{\S\S\xspace#3#1#4 to~#5#2#6}
\crefmultiformat{section}{\S\S\xspace#2#1#3}{ and~#2#1#3}{, #2#1#3}{ and~#2#1#3}

\BeforeBeginEnvironment{wrapfigure}{\setlength{\intextsep}{3pt}}

\newcolumntype{R}[1]{>{\raggedleft\arraybackslash}p{#1}}

\newcommand{\squishitemize}{
  \begin{list}{$\bullet$}{
    \setlength{\itemsep}{0pt}
    \setlength{\parsep}{3pt}
    \setlength{\topsep}{3pt}
    \setlength{\partopsep}{0pt}
    \setlength{\leftmargin}{1.95em}
    \setlength{\labelwidth}{1.5em}
    \setlength{\labelsep}{0.5em} } }

\newcounter{Lcount}
\newcommand{\squishlist}{
  \begin{list}{\arabic{Lcount}. }{
    \usecounter{Lcount}
    \setlength{\itemsep}{0pt}
    \setlength{\parsep}{3pt}
    \setlength{\topsep}{3pt}
    \setlength{\partopsep}{0pt}
    \setlength{\leftmargin}{2em}
    \setlength{\labelwidth}{1.5em}
    \setlength{\labelsep}{0.5em} } }

\newcommand{\squishend}{\end{list}}

\newcommand{\etal}{\textit{et al.}\xspace}

\newcommand{\eg}{\textit{e.g.},\xspace}

\renewcommand{\emph}[1]{\textit{#1}}

\renewcommand\footnotetextcopyrightpermission[1]{}  
\acmConference{}{}{}
\acmYear{}

\usepackage{xargs}
\usepackage{tikz}
\usepackage{xcolor}
\usepackage{enumitem}

\tikzset{
  PrincipleTag/.style={
    draw,
    font=\small\sffamily,
    align=center
  },
  PrincipleLargeTag/.style={
    draw,
    minimum width=1cm,
    minimum height=1cm,
    font=\small\sffamily,
    align=center
  },
  StructureBox/.style   ={PrincipleTag,fill=blue!15},        
  SemanticsBox/.style   ={PrincipleTag,fill=yellow!20},      
  DistributionBox/.style  ={PrincipleTag,fill=teal!20},      
  EfficiencyBox/.style  ={PrincipleTag,fill=orange!20},      
  PlanningBox/.style    ={PrincipleTag,fill=green!20},       
  OperabilityBox/.style ={PrincipleTag,fill=cyan!20},        
  ReliabilityBox/.style ={PrincipleTag,fill=red!15},         
  SecurityBox/.style    ={PrincipleTag,fill=violet!20},      
  StructureLargeBox/.style   ={PrincipleLargeTag,fill=blue!15},        
  SemanticsLargeBox/.style   ={PrincipleLargeTag,fill=yellow!20},      
  DistributionLargeBox/.style  ={PrincipleLargeTag,fill=teal!20},      
  EfficiencyLargeBox/.style  ={PrincipleLargeTag,fill=orange!20},      
  PlanningLargeBox/.style    ={PrincipleLargeTag,fill=green!20},       
  OperabilityLargeBox/.style ={PrincipleLargeTag,fill=cyan!20},        
  ReliabilityLargeBox/.style ={PrincipleLargeTag,fill=red!15},         
  SecurityLargeBox/.style    ={PrincipleLargeTag,fill=violet!20},      
}

\newcommand{\StructureTag}[1]{\ptbox[StructureBox]{#1}}
\newcommand{\SemanticsTag}[1]{\ptbox[SemanticsBox]{#1}}

\newcommand{\EfficiencyTag}[1]{\ptbox[EfficiencyBox]{#1}}
\newcommand{\PlanningTag}[1]{\ptbox[PlanningBox]{#1}}
\newcommand{\OperabilityTag}[1]{\ptbox[OperabilityBox]{#1}}

\newcommand{\ptbox}[2][]{%
  \tikz[baseline=(pt.base)]\node[PrincipleTag,#1](pt){#2};%
}

\newcommandx{\principle}[3][3=]{%
  \item[\textbf{#1}.] #2%
  \ifx\\#3\\%
  \else\\\textit{Example:} #3%
  \fi
}

\begin{document}

\title{\paperTitle}
\author{Joy Arulraj}
\affiliation{%
  \institution{Georgia Institute of Technology}
  \city{Atlanta}
  \state{GA}
  \country{USA}
}
\email{arulraj@gatech.edu}

\begin{abstract}
System design is often taught through domain-specific solutions specific to particular domains, such as databases, operating systems, or computer architecture, each with its own methods and vocabulary. 
While this diversity is a strength, it can obscure cross-cutting principles that recur across domains. 
This paper proposes a preliminary “periodic table” of system design principles distilled from several domains in computer systems.
The goal is a shared, concise vocabulary that helps students, researchers, and practitioners reason about structure and trade-offs, compare designs across domains, and communicate choices more clearly.
For supporting materials and updates, please refer to the repository at: \href{https://github.com/jarulraj/periodic-table}{https://github.com/jarulraj/periodic-table}.
\end{abstract}

\maketitle
\vspace{-1.15em}  

\section{Introduction}

One of the rewards of working in computer systems is the field's sheer diversity, spanning operating systems, databases, computer architecture, distributed systems, programming languages, networking, and more, each with a rich history.
For newcomers, it can be challenging to spot connections across different domains due to the diversity of traditions and vocabularies: the same design principle may appear in different guises across domains.

For example, consider the classic paper on database isolation levels by Jim Gray \etal.~\cite{gray75}.
It offers a careful account of concurrency-control mechanisms and the trade-offs between correctness and performance.
Yet without prior exposure to similar issues in operating systems or computer architecture, the ideas can appear narrowly “about databases.”
In reality, the same design principle, "relaxation of consistency," reappears across systems in different guises, from weakly ordered memory hierarchies to eventual-consistency protocols.
When each community uses its own terms and exemplars, newcomers may find it difficult to recognize the underlying design principles.
This fragmentation increases cognitive overhead, as the same trade-off must be relearned in each context.

This is a broader pattern: systems research is rich in practical insight but lighter on shared conceptual scaffolding.
Across domains, similar challenges recur, managing concurrency, ensuring consistency, and adapting to change, while the framing and vocabulary often differ.
As a result, deep connections between seemingly disparate domains can remain relatively obscure.

This article is a small step toward bridging those gaps.
Borrowing Mendeleev's metaphor, we present a "periodic table" of recurring system design principles.
The goal is not a rigid taxonomy but a working vocabulary: a concise way to annotate papers, lectures, and design documents with the fundamental principles they employ.
The aim is to surface structure that already exists in computer systems, so that students can form a more coherent mental map, researchers can situate contributions with precision, and practitioners can discuss design choices across domains with greater clarity.

\section{Methodology}

\begin{table*}[h]
\centering
\begin{tabular}{@{}ll@{}}
\toprule
\textbf{Group} & \textbf{Focus of the Group} \\
\midrule
\textbf{Structure}    & How to carve and connect parts with clear boundaries and extension points. \\
\textbf{Efficiency}   & Do less work, or do it cheaper by focusing effort where it pays. \\
\textbf{Semantics}    & Specify behavior and interfaces precisely. \\
\textbf{Distribution} & Coordinate work and data across distributed architectures. \\
\textbf{Planning}     & Select plans automatically from goals, costs, and constraints. \\
\textbf{Operability}  & Observe, adapt, and evolve running systems with minimal disruption. \\
\textbf{Reliability}  & Stay correct under faults, concurrency, and partial failure. \\
\textbf{Security}     & Bound authority and enforce isolation to preserve safety and integrity. \\
\bottomrule
\end{tabular}
\caption{Thematic groups of design principles}
\label{tab:principle-groups}
\end{table*}

We identified principles by going over 100+ influential papers across operating systems, computer architecture, databases, networking, programming languages, security, and other domains in computer systems. 
These papers were chosen for historical significance and ongoing relevance, such as classic papers on concurrency control~\cite{gray75} and consensus~\cite{lamport}, and more recent work on using machine learning inside systems~\cite{jimenez01} and designing systems for the cloud~\cite{chase01}.
For each paper we asked: what is the underlying high-level design principle?
Across domains, independent systems often converged not on mechanisms but on shared design principles: for example, relaxing consistency to improve performance or lifting abstractions to enhance usability. 
To qualify as a system design principle, it must satisfy two conditions:

\begin{enumerate}
  \item \textbf{Abstract.}
  The principle must be independent of specific technologies or implementations. 

  \item \textbf{Generality.}
  The principle must show up across different domains (\eg database systems, operating systems, programming languages).
\end{enumerate}

This analysis does not aim to catalogue every principle, but to surface many of those with lasting, general-purpose value.

\section{The Design Principle Table}

We have curated a structured set of 40+ general-purpose design principles distilled from the systems literature. 
As shown in \cref{tab:principle-groups}, they are organised into thematic groups that mirror familiar axes of system design.

Each principle is tagged with a short symbol (e.g., \texttt{Co} for composability, \texttt{Op} for optimistic design) for quick reference. 
We emphasise \emph{design intent} rather than prescribing mechanisms: instead of “use this locking protocol” or “optimise this query plan,” the principles state aims such as “preserve correctness under concurrency” or “prioritise the common case,” leaving concrete realisations to specific domains.

\begin{figure*}
\begin{tikzpicture}[font=\sffamily]

\def\rowgap{1.3} 

\foreach[count=\i] \sym in {Si,Mo,Co,Ex,Pm,Gr}{
  \node[StructureLargeBox] at ({0},-\i*\rowgap) {\sym};
}

\foreach[count=\i] \sym in {Sc,Rc,Wv,Cc,Bo,Ha,Op,La}{
  \node[EfficiencyLargeBox] at ({1*1.3},-\i*\rowgap) {\sym};
}

\foreach[count=\i] \sym in {Al,Lu,Se,Fs,Ig}{
  \node[SemanticsLargeBox] at ({2*1.3},-\i*\rowgap) {\sym};
}

\foreach[count=\i] \sym in {Lt,Dc,Fp,Lo}{
  \node[DistributionLargeBox] at ({3*1.3},-\i*\rowgap) {\sym};
}

\foreach[count=\i] \sym in {Ep,Cm,Cp,Gd,Bb,Ah}{
  \node[PlanningLargeBox] at ({4*1.3},-\i*\rowgap) {\sym};
}

\foreach[count=\i] \sym in {Ad,Ec,Wa,Au,Ho,Ev}{
  \node[OperabilityLargeBox] at ({5*1.3},-\i*\rowgap) {\sym};
}

\foreach[count=\i] \sym in {Ft,Is,At,Cr}{
  \node[ReliabilityLargeBox] at ({6*1.3},-\i*\rowgap) {\sym};
}

\foreach[count=\i] \sym in {Sy,Ac,Lp,Tq,Cf,Sa}{
  \node[SecurityLargeBox] at ({7*1.3},-\i*\rowgap) {\sym};
}

\node[above=2pt] at (0,0)     {\textbf{Str}};
\node[above=2pt] at (1*1.3,0) {\textbf{Eff}};
\node[above=2pt] at (2*1.3,0) {\textbf{Sem}};
\node[above=2pt] at (3*1.3,0) {\textbf{Dist}};
\node[above=2pt] at (4*1.3,0) {\textbf{Plan}};
\node[above=2pt] at (5*1.3,0) {\textbf{Oper}};
\node[above=2pt] at (6*1.3,0) {\textbf{Rel}};
\node[above=2pt] at (7*1.3,0) {\textbf{Sec}};

\end{tikzpicture}
\end{figure*}

\subsection{Group 1: Structure}
\label{subsec:group1}

\begin{description}

\principle{Si – Simplicity}{
Choose the simplest system design that meets current needs; resist  complexity, such as additional layers, services, or generality added "just in case", until evidence shows benefit. Avoid premature \emph{architectural} optimisation of the system~\cite{knuth74}.
}

\principle{Mo – Modularity}{
Partition the system into cohesive units with minimal interfaces, so that each unit can be reasoned about, replaced, or evolved independently. 
This principle focuses on decomposition: choosing boundaries to favor  clear separation of concerns so that each responsibility sits in one module.
}
[The OSI model decomposes communication into standardised layers with well-defined boundaries that permit independent development and substitution~\cite{zimmermann80}.]

\principle{Co – Composability}{
Design components that can be safely and flexibly recombined; rely on explicit contracts and type-constrained interfaces so that every legal composition remains correct, letting components be assembled like interchangeable bricks.
Unlike modularity, this principle focuses on re-composition: making sure the components can be combined safely and flexibly.
}
[Unix programs (e.g., \texttt{grep}, \texttt{sort}, \texttt{uniq}) read from stdin and write to stdout, letting the user compose complex text processing pipelines~\cite{ritchie74}]

\principle{Ex – Extensibility}{
Design systems to allow safe user-defined extensions, such as plug-ins, without requiring changes to the system core.
When extensions come from untrusted parties, isolate them through sandboxing to preserve safety.
}
[Unix also illustrates extensibility: new programs can be added by the user without kernel changes~\cite{ritchie74}.]

\principle{Pm – Policy/Mechanism Separation}{
Separate what should be done (policy) from how it is carried out (mechanism) by exposing a common interface through which multiple policies can plug into the same mechanism.
}
[Hydra has a kernel of generic mechanisms (scheduling, paging, protection) and moved resource-allocation policies to user-level modules~\cite{levin75}.]

\principle{Gr – Generalized Design}{
Design a single core with explicit variation points like types, knobs, or plug-ins, so that it can serve many use cases without duplication, but specialise when doing so yields meaningful gains in performance, accuracy, or clarity.
}
[The C++ Standard Template Library is a collection of containers, iterators, and algorithms parameterized by templates~\cite{stepanov94}.
Postgres allows users to add types and operators to the core database system~\cite{stonebraker86}.
]

\principle{Pd – Probabilistic Design}{
Introduce controlled randomness to gain efficiency, scalability, or simplicity while accepting a small, quantified risk of error or loss.
}
[Routers treat queue length as a probability signal: as the queue grows, they drop incoming packets with increasing probability, proactively signalling congestion~\cite{floyd93}.]

\end{description}

\subsection{Group 2: Efficiency}
\label{subsec:group2}

\begin{description}
\principle{Sc – Scalability}{
Design the system to handle growth in data, traffic, or nodes with near-linear cost or latency.
}
[MapReduce scales across nodes by dividing work into parallel tasks and aggregating results with minimal coordination~\cite{dean04}.]

\principle{Rc – Reuse of Computation}{
Avoid redundant work by caching, materializing intermediate results (e.g., indexes), or incrementally updating outputs across repeated or slightly modified inputs, saving computation.
}
[A B+tree reuses its sorted key order: lookups follow the existing search path instead of rescanning the entire data set each time, thereby reusing computation~\cite{bayer72}.]

\principle{Wv – Work Avoidance}{
Skip computation that would not alter externally observable results.
Examples include lazy evaluation and predicate short-circuiting.
}
[Lazy evaluation defers work until a value is demanded, eliminating useless computation~\cite{hughes90}.]

\principle{Cc – Common-Case Specialization}{
Detect the execution paths or data items that dominate run-time ("hot spots") and create a streamlined fast path just for them, while a slower, general path still handles every case correctly.
}
[Caching the target method for the receiver class on the first call, so that subsequent calls on that common receiver hit the fast path; uncommon classes fall back to the full method-lookup routine~\cite{chambers89}.]

\principle{Bo – Bottleneck-Oriented Optimisation}{
Profile end-to-end performance, locate the tightest resource constraint, and focus improvement effort there until another stage becomes the limiter.
}
[Rare 99\textsuperscript{th}-percentile stragglers bottleneck latency, and replicated requests help cut tail response times~\cite{dean13}.]

\principle{Ha – Hardware-Aware Design}{
Shape algorithms and data structures to the latency, bandwidth, parallelism, and persistence properties of underlying hardware (e.g., cache hierarchy, NUMA, SSDs, GPUs).
}
[BLAS defines cache- and vector-tuned kernels so linear-algebra code exploits hardware efficiently~\cite{lawson79blas}.]

\principle{Op – Optimistic Design}{
Proceed as if the common case will succeed, skipping coordination, and rely on a (possibly expensive) recovery path only when that assumption proves wrong.}
[Optimistic Concurrency Control runs transactions lock-free, then validates at commit and rolls back only when a conflict is detected~\cite{kung81}.]

\principle{La – Learned Approximation}{
Replace hand-crafted algorithms with models trained on data, trading bounded inaccuracy for efficiency or flexibility.
}
[The perceptron branch predictor learns weights online to forecast branch outcomes, outperforming fixed two-bit counters without enlarging the table~\cite{jimenez01}.
]
\end{description}

\subsection{Group 3: Semantics}
\label{subsec:group3}

\begin{description}
\principle{Al – Abstraction Lifting}{
Wrap low-level operations behind a higher-level interface or domain-specific language that expresses \textit{intent} rather than steps.
This enables internal optimization and also allows a single definition to target diverse back-ends.}
[SQL queries declare the result to retrieve; the DBMS chooses access paths, join orders, and physical operators automatically~\cite{selinger79}.]

\principle{Lu – Language Homogeneity}{
Adopt a single, well-specified intermediate representation (or language) across core components and extensions so semantics align, tools compose, and cross-layer optimisations and reuse happen with minimal effort.
}
[
LLVM exposes a typed, SSA-based IR that many front ends target and many back ends share, enabling cross-language optimisation and reuse of the same middle-end passes~\cite{lattner04}.
]

\principle{Se – Semantically Explicit Interfaces}{
Specify an interface precisely (covering effect visibility, ordering, durability, etc.) so that users can reason about a call's true externally observable state without guessing about hidden buffering or replication.
}
[SQL isolation levels specify precise anomaly semantics and make visibility guarantees explicit~\cite{berenson95}.
]

\principle{Fs – Formal Specification}{
Describe system behaviour using mathematical models or logic to support rigorous reasoning, verification, or synthesis.
Mechanisms for realizing this principle include temporal logic, state machines, and other formalisms that make system properties analyzable.
}
[TLA\texttt{+} shows how to specify and check systems using logic and set theory to catch design errors before coding.~\cite{Lamport02}.]

\principle{Ig – Invariant‑Guided Transformation}{
Use formally stated invariants to drive safe refactoring, optimisation, or reconfiguration.
}
[
In compilers, SSA treats "one definition per name"  as an IR invariant; passes rewrite code while preserving semantics and then re-establish SSA~\cite{cytron91ssa}. 
In query optimisers, relational-algebra equivalences (e.g., selection/projection pushdown) preserve result semantics~\cite{selinger79}.
]

\end{description}

\subsection{Group 4: Distribution}
\label{subsec:group4}

\begin{description}
\principle{Lt – Location Transparency}{
Hide the physical whereabouts of resources so clients interact via uniform names or handles.
}
[Programs can call remote procedures as if they were local, masking host location~\cite{birrell84}.]

\principle{Dc – Decentralised Control}{
Distribute decision‑making among many nodes to avoid single points of failure or bottlenecks.
}
[Dynamo partitions data via consistent hashing and uses gossip-based membership, avoiding any central coordinator~\cite{decandia07}.]

\principle{Fp – Function Placement}{
Place functionality where the necessary context and resources exist to achieve correctness and efficiency, avoiding redundant work elsewhere.
}
[
The end-to-end argument shows that functions like reliability checks achieve correctness only at the endpoints~\cite{saltzer84}.
]

\principle{Lo – Locality of Reference}{
Place related data and operations close together in time and space to preserve access patterns and minimize separation between computation and state.
}
[The working-set model formalises temporal locality to keep hot pages in memory~\cite{denning68}.]

\end{description}

\subsection{Group 5: Planning}
\label{subsec:group5}

\begin{description}

\principle{Ep – Equivalence-based Planning}{
Apply algebraic/logic rewrite rules over a common IR that preserve semantic equivalence; defer final choice to later cost/constraint stages.
}
[
Starburst's rule-based rewrite system applies relational equivalences (e.g., predicate pushdown) to generate logically equivalent queries~\cite{pirahesh92}.
]

\principle{Cm – Cost-based Planning}{
When a system must choose among alternative designs, configurations, or execution strategies, use a cost model to guide the search toward low-cost solutions (energy, money, e.t.c.) without needing to enumerate the full space.
}
[The Selinger query optimizer selects the lowest-cost plan under a cost model~\cite{selinger79}.
]

\principle{Cp – Constraint-based Planning}{
Encode decisions and hard or soft constraints and rely on a solver (ILP/SMT e.t.c.) to find a feasible or optimal assignment.
}
[Quincy formulates cluster scheduling as a min-cost flow with locality and fairness constraints and solves it to obtain an assignment~\cite{isard09}.]

\principle{Gd – Goal-Directed Planning}{
Accept a declarative description of the desired end-state and automatically synthesise a concrete sequence of operations to reach it, shielding the user from implementation details.
}
[The Cascades query optimizer turns an SQL query (the goal) into an executable plan via rule-based transformation and cost-guided search~\cite{graefe95}.]

\principle{Bb – Black-Box Tuning}{
When analytic cost models are not available, search the plan/configuration space by \emph{measuring} candidates on the target system, iteratively choosing better ones (e.g., heuristic or Bayesian search), and caching the winner.
}
[ATLAS empirically times candidate BLAS kernel configurations on the target CPU and fixes the best-performing parameters, without an analytic cost model~\cite{whaley98}.
]

\principle{Ah – Advisory Hinting}{
Provide non-binding hints that systems may exploit to improve performance, without changing correctness or requiring enforcement.}
[Lampson advocates optional "hints" that help performance but must not affect correctness if ignored~\cite{lampson83}.]

\end{description}

\subsection{Group 6: Operability}
\label{subsec:group6}

\begin{description}

\principle{Ad – Adaptive Processing}{
Monitor runtime conditions and automatically adjust parameters or strategy.
}
[Eddies continuously reorder query operators at runtime based on feedback, adapting without stopping execution~\cite{avnur00}.]

\principle{Ec – Elasticity}{
Automatically adjust resource allocation in response to shifting demand and cost goals
Examples include predictive autoscaling and load shaping.
}
[Chase \emph{et~al.} dynamically provision servers based on load and utility, exemplifying elastic resource management~\cite{chase01}.]

\principle{Wa – Workload-Aware Optimisation}{
Continuously observe workload shape (skew, locality, access frequency e.t.c.), and adapt data layouts, algorithm choices, or resource allocations to match current patterns.
}
[Database "cracking" incrementally reorganises column data based on query predicates, adapting the data layout continuously to the observed workload~\cite{idreos07}.]

\principle{Au – Automation and Autonomy}{
Let the system perform routine or reactive tasks without human intervention, often by learning from traces or user-provided examples.
}
[AutoAdmin automatically recommends indexes/materialized views from workload traces~\cite{chaudhuri97}.
Programming-by-example systems automate tasks by generalizing from a few user-provided examples~\cite{lieberman01}.
]

\principle{Ho – Human Observability}{
Expose internal state of the system, like metrics, traces, plans, to make the system intentionally transparent; that transparency improves observability, debugging, introspection, and control.
}
[Paxson's end-to-end Internet packet dynamics analysis demonstrates how rich measurement and tracing enable informed debugging and tuning.~\cite{Paxson99}]

\principle{Ev – Evolvability}{
Design so the system can change with minimal downtime or rewrites and do so without breaking external contracts or observable behaviour for existing clients.
Unlike extensibility that lets outsiders add new behavior via defined hook points without touching the core, evolvability lets the system's internals change over time without breaking existing external contracts.
}
[Parnas presents how a modular design makes system easier to extend without disruptive rewrites~\cite{Parnas79}.]

\end{description}

\subsection{Group 7: Reliability}
\label{subsec:group7}

\begin{description}

\principle{Ft – Fault Tolerance}{
Design the system to continue operating, perhaps in degraded form, despite component failures.
}
[Gray's analysis of why computers stop shows that replication and automatic restart let services keep running through hardware and software faults~\cite{gray86}.]

\principle{Is – Isolation for Correctness}{
Prevent unintended interference among components so local reasoning remains valid.
}
[Two-phase row-level locking stops one transaction from reading or overwriting another’s uncommitted data, preserving isolation guarantees~\cite{gray93}.]

\principle{At – Atomic Execution}{
Group multiple operations so they appear indivisible, either all take effect or none do.
}
[With Transactional Memory, memory operations inside a transaction speculatively execute, then commit atomically; if any conflict or fault occurs, the entire block aborts and leaves no partial state~\cite{herlihy93}.
]

\principle{Cr – Consistency Relaxation}{
Deliberately relax strong consistency or ordering constraints, but only within documented bounds, to improve performance, availability, or concurrency.
}
[Bayou lets mobile clients update replicas while disconnected, guaranteeing eventual convergence when replicas reconnect, trading strict consistency for offline availability~\cite{petersen97}.]

\end{description}

\subsection{Group 8: Security}
\label{subsec:group8}

\begin{description}
\principle{Sy – Security via Isolation}{
Enforce strong boundaries so faults or hostile code cannot affect other components.
}
[
A correct virtual machine monitor presents each guest with a complete, isolated machine and intercepts privileged operations, preventing one guest from compromising others or the host~\cite{popek74}.]

\principle{Ac – Access Control and Auditing}{
Define permissions \emph{and} log every access for accountability.
}
[Lampson’s taxonomy of access-control lists, capabilities, and audit trails  underpins modern security mechanisms~\cite{lampson71}.]

\principle{Lp – Least Privilege}{
Grant only minimal authority needed for a task, shrinking the blast radius.
}
[The post-mortem on the 1988 Internet Worm shows how excess privilege let the worm spread and spurred widespread adoption of least-privilege daemons~\cite{morris89}.]

\principle{Tq – Trust via Quorum}{
Rely on agreement from multiple, independent participants rather than a single authority.
}
[Paxos algorithm replicates state across a majority quorum so the service stays correct even if minority nodes crash or act maliciously~\cite{lamport98}.]

\principle{Cf – Conservative Defaults}{
Ship with restrictive, safe settings; let experts opt-in to riskier, faster modes.
}
[With a "default no-access" policy, every protection mechanism should allow access only when explicitly granted~\cite{saltzer75}.]

\principle{Sa – Safety by Construction}{
Structure code or data so entire classes of errors become impossible rather than merely detected.
}
[Rust’s ownership and borrow checker prevent data races and dangling pointers at compile time~\cite{matsakis14}.]

\end{description}

\section{Case Study}
\label{sec:case-study}

To illustrate how multiple design principles intersect in practice, consider the mapping from logical to physical operator plans in a relational database system.
The database system translates declarative intent into executable steps (\StructureTag{Policy/Mechanism Separation}).
SQL expresses the "what" (\SemanticsTag{Abstraction Lifting}) with precise semantics (\SemanticsTag{Semantically Explicit Interfaces}). 
The optimizer first rewrites the query using algebraic equivalences (\PlanningTag{Equivalence-Based Planning}). 
It then chooses concrete physical operators using a cost model  (\PlanningTag{Cost-based Planning}).
Physical operators are often optimized for underlying hardware features (\EfficiencyTag{Hardware-Aware Design}).

Predicate-pushdown illustrates \EfficiencyTag{Work Avoidance}, while indexes enable \EfficiencyTag{Reuse of Computation}. 
\PlanningTag{Advisory Hints} can guide the optimizer, and newer database systems add runtime re-optimization (\OperabilityTag{Adaptive Processing}), learned models (\EfficiencyTag{Learned Approximation}), and sampling (\StructureTag{Probabilistic Design}).
Thus, logical-to-physical operator mapping in database systems exemplifies how several design principles come together to efficiently process declarative SQL queries.

\section{Limitations}
\label{sec:limitations}

Any attempt to organise a field as broad as computer systems involves trade-offs.
This table is not a checklist or a universal theory; it is a shared vocabulary that highlights recurring principles and encourages structural reflection.
That said, there are several limitations:

\begin{itemize}
  \item \textbf{Orthogonality.}
  Principles can overlap, reinforce, or partially conflict; design is about balancing such tensions.
  \item \textbf{Subjectivity and granularity.}
  Deriving and mapping principles involves judgement; boundaries are fuzzy and different readers may tag the same system differently or interpret the same principle differently.
  \item \textbf{Not a formal taxonomy.}
  This is not a complete or minimal set of design principles.
  There is no attempt to derive the principles from a minimal core.
\end{itemize}

Ultimately, this table is a means to help students see recurring design principles more clearly, to assist system designers in communicating tradeoffs more precisely, and to help researchers recognize where their ideas fit into the broader landscape of system design.
%

\section{Conclusion}
\label{sec:conclusion}

System design spans diverse domains and vocabularies, which can make shared discussion harder. 
We inherit mechanisms, study trade-offs, and build intuitions, yet concise terms for the underlying ideas are not always available.
The “periodic table” of design principles offered here aims to provide a modest common language, naming recurring ideas so they are easier to teach, compare, and build upon.

\bibliographystyle{ACM-Reference-Format}
\bibliography{refs}


\begin{thebibliography}{48}


\ifx \showCODEN    \undefined \def \showCODEN     #1{\unskip}     \fi
\ifx \showISBNx    \undefined \def \showISBNx     #1{\unskip}     \fi
\ifx \showISBNxiii \undefined \def \showISBNxiii  #1{\unskip}     \fi
\ifx \showISSN     \undefined \def \showISSN      #1{\unskip}     \fi
\ifx \showLCCN     \undefined \def \showLCCN      #1{\unskip}     \fi
\ifx \shownote     \undefined \def \shownote      #1{#1}          \fi
\ifx \showarticletitle \undefined \def \showarticletitle #1{#1}   \fi
\ifx \showURL      \undefined \def \showURL       {\relax}        \fi
\providecommand\bibfield[2]{#2}
\providecommand\bibinfo[2]{#2}
\providecommand\natexlab[1]{#1}
\providecommand\showeprint[2][]{arXiv:#2}

\bibitem[Avnur and Hellerstein(2000)]%
        {avnur00}
\bibfield{author}{\bibinfo{person}{Ron Avnur} {and} \bibinfo{person}{Joseph~M. Hellerstein}.} \bibinfo{year}{2000}\natexlab{}.
\newblock \showarticletitle{Eddies: Continuously Adaptive Query Processing}. In \bibinfo{booktitle}{\emph{SIGMOD}}.
\newblock


\bibitem[Bayer and McCreight(1972)]%
        {bayer72}
\bibfield{author}{\bibinfo{person}{Rudolf Bayer} {and} \bibinfo{person}{Edward McCreight}.} \bibinfo{year}{1972}\natexlab{}.
\newblock \showarticletitle{Organization and Maintenance of Large Ordered Indexes}.
\newblock \bibinfo{journal}{\emph{Acta Informatica}} (\bibinfo{year}{1972}).
\newblock


\bibitem[Berenson et~al\mbox{.}(1995)]%
        {berenson95}
\bibfield{author}{\bibinfo{person}{Hal Berenson}, \bibinfo{person}{Philip~A. Bernstein}, \bibinfo{person}{Jim Gray}, \bibinfo{person}{Jim Melton}, \bibinfo{person}{Elizabeth~J. O'Neil}, {and} \bibinfo{person}{Patrick~E. O'Neil}.} \bibinfo{year}{1995}\natexlab{}.
\newblock \showarticletitle{A Critique of {ANSI} {SQL} Isolation Levels}. In \bibinfo{booktitle}{\emph{SIGMOD}}.
\newblock


\bibitem[Birrell and Nelson(1984)]%
        {birrell84}
\bibfield{author}{\bibinfo{person}{Andrew~D. Birrell} {and} \bibinfo{person}{Bruce~J. Nelson}.} \bibinfo{year}{1984}\natexlab{}.
\newblock \showarticletitle{Implementing Remote Procedure Calls}.
\newblock \bibinfo{journal}{\emph{ACM TOCS}} (\bibinfo{year}{1984}).
\newblock


\bibitem[Chambers and Ungar(1989)]%
        {chambers89}
\bibfield{author}{\bibinfo{person}{Craig Chambers} {and} \bibinfo{person}{David Ungar}.} \bibinfo{year}{1989}\natexlab{}.
\newblock \showarticletitle{Customization: Optimizing Compiler Technology for {SELF}, a Dynamically Typed Object-Oriented Programming Language}. In \bibinfo{booktitle}{\emph{PLDI}}.
\newblock


\bibitem[Chase et~al\mbox{.}(2001)]%
        {chase01}
\bibfield{author}{\bibinfo{person}{Jeffrey~S. Chase}, \bibinfo{person}{Darrell~C. Anderson}, \bibinfo{person}{Prachi~N. Thakar}, \bibinfo{person}{Amin Vahdat}, {and} \bibinfo{person}{Ronald~P. Doyle}.} \bibinfo{year}{2001}\natexlab{}.
\newblock \showarticletitle{Managing Energy and Server Resources in Hosting Centers}. In \bibinfo{booktitle}{\emph{SOSP}}. \bibinfo{address}{Banff, Canada}.
\newblock


\bibitem[Chaudhuri and Narasayya(1997)]%
        {chaudhuri97}
\bibfield{author}{\bibinfo{person}{Surajit Chaudhuri} {and} \bibinfo{person}{Vivek~R. Narasayya}.} \bibinfo{year}{1997}\natexlab{}.
\newblock \showarticletitle{An Efficient, Cost-Driven Index Selection Tool for Microsoft SQL Server}. In \bibinfo{booktitle}{\emph{VLDB}}.
\newblock


\bibitem[Cytron et~al\mbox{.}(1991)]%
        {cytron91ssa}
\bibfield{author}{\bibinfo{person}{Ron Cytron}, \bibinfo{person}{Jeanne Ferrante}, \bibinfo{person}{Barry~K. Rosen}, \bibinfo{person}{Mark~N. Wegman}, {and} \bibinfo{person}{F.~Kenneth Zadeck}.} \bibinfo{year}{1991}\natexlab{}.
\newblock \showarticletitle{Efficiently Computing Static Single Assignment Form and the Control Dependence Graph}.
\newblock \bibinfo{journal}{\emph{ACM Transactions on Programming Languages and Systems}} (\bibinfo{year}{1991}).
\newblock


\bibitem[Dean and Barroso(2013)]%
        {dean13}
\bibfield{author}{\bibinfo{person}{Jeff Dean} {and} \bibinfo{person}{Luiz~Andr{\'e} Barroso}.} \bibinfo{year}{2013}\natexlab{}.
\newblock \showarticletitle{The Tail at Scale}.
\newblock \bibinfo{journal}{\emph{Commun. ACM}} (\bibinfo{year}{2013}).
\newblock


\bibitem[Dean and Ghemawat(2004)]%
        {dean04}
\bibfield{author}{\bibinfo{person}{Jeffrey Dean} {and} \bibinfo{person}{Sanjay Ghemawat}.} \bibinfo{year}{2004}\natexlab{}.
\newblock \showarticletitle{MapReduce: Simplified Data Processing on Large Clusters}. In \bibinfo{booktitle}{\emph{OSDI}}.
\newblock


\bibitem[Denning(1968)]%
        {denning68}
\bibfield{author}{\bibinfo{person}{Peter~J. Denning}.} \bibinfo{year}{1968}\natexlab{}.
\newblock \showarticletitle{The Working Set Model for Program Behavior}.
\newblock \bibinfo{journal}{\emph{Commun. ACM}} (\bibinfo{year}{1968}).
\newblock


\bibitem[et~al.(2007)]%
        {decandia07}
\bibfield{author}{\bibinfo{person}{Giuseppe~DeCandia et al.}} \bibinfo{year}{2007}\natexlab{}.
\newblock \showarticletitle{Dynamo: Amazon's Highly Available Key-Value Store}. In \bibinfo{booktitle}{\emph{SOSP}}.
\newblock


\bibitem[Floyd and Jacobson(1993)]%
        {floyd93}
\bibfield{author}{\bibinfo{person}{Sally Floyd} {and} \bibinfo{person}{Van Jacobson}.} \bibinfo{year}{1993}\natexlab{}.
\newblock \showarticletitle{Random Early Detection Gateways for Congestion Avoidance}. In \bibinfo{booktitle}{\emph{SIGCOMM}}.
\newblock


\bibitem[Graefe(1995)]%
        {graefe95}
\bibfield{author}{\bibinfo{person}{Goetz Graefe}.} \bibinfo{year}{1995}\natexlab{}.
\newblock \bibinfo{booktitle}{\emph{The Cascades Framework for Query Optimisation}}.
\newblock \bibinfo{type}{{T}echnical {R}eport} HPL-95-18. \bibinfo{institution}{Hewlett–Packard Labs}.
\newblock


\bibitem[Gray(1986)]%
        {gray86}
\bibfield{author}{\bibinfo{person}{Jim Gray}.} \bibinfo{year}{1986}\natexlab{}.
\newblock \showarticletitle{Why Do Computers Stop and What Can Be Done About It?}
\newblock \bibinfo{journal}{\emph{Tandem Technical Report}} (\bibinfo{year}{1986}).
\newblock


\bibitem[Gray and Reuter(1993)]%
        {gray93}
\bibfield{author}{\bibinfo{person}{Jim Gray} {and} \bibinfo{person}{Andreas Reuter}.} \bibinfo{year}{1993}\natexlab{}.
\newblock \bibinfo{booktitle}{\emph{Transaction Processing: Concepts and Techniques}}.
\newblock \bibinfo{publisher}{Morgan Kaufmann}.
\newblock


\bibitem[Gray et~al\mbox{.}(1975)]%
        {gray75}
\bibfield{author}{\bibinfo{person}{J.~N. Gray}, \bibinfo{person}{R.~A. Lorie}, {and} \bibinfo{person}{G.~R. Putzolu}.} \bibinfo{year}{1975}\natexlab{}.
\newblock \showarticletitle{Granularity of locks in a shared data base}. In \bibinfo{booktitle}{\emph{VLDB}}.
\newblock


\bibitem[Herlihy and Moss(1993)]%
        {herlihy93}
\bibfield{author}{\bibinfo{person}{Maurice Herlihy} {and} \bibinfo{person}{J.~Eliot~B. Moss}.} \bibinfo{year}{1993}\natexlab{}.
\newblock \showarticletitle{Transactional Memory: Architectural Support for Lock-Free Data Structures}. In \bibinfo{booktitle}{\emph{ISCA}}. \bibinfo{publisher}{ACM}.
\newblock


\bibitem[Hughes(1990)]%
        {hughes90}
\bibfield{author}{\bibinfo{person}{John Hughes}.} \bibinfo{year}{1990}\natexlab{}.
\newblock \showarticletitle{Why Functional Programming Matters}.
\newblock In \bibinfo{booktitle}{\emph{Research Topics in Functional Programming}}. \bibinfo{publisher}{Addison-Wesley}.
\newblock


\bibitem[Idreos et~al\mbox{.}(2007)]%
        {idreos07}
\bibfield{author}{\bibinfo{person}{Stratos Idreos}, \bibinfo{person}{Martin~L. Kersten}, {and} \bibinfo{person}{Stefan Manegold}.} \bibinfo{year}{2007}\natexlab{}.
\newblock \showarticletitle{Database Cracking}. In \bibinfo{booktitle}{\emph{CIDR}}.
\newblock


\bibitem[Isard et~al\mbox{.}(2009)]%
        {isard09}
\bibfield{author}{\bibinfo{person}{Michael Isard}, \bibinfo{person}{Vijayan Prabhakaran}, \bibinfo{person}{Jon Currey}, \bibinfo{person}{Udi Wieder}, \bibinfo{person}{Christopher Schmitz}, {and} \bibinfo{person}{Alexander Totok}.} \bibinfo{year}{2009}\natexlab{}.
\newblock \showarticletitle{Quincy: Fair Scheduling for Distributed Computing Clusters}. In \bibinfo{booktitle}{\emph{SOSP}}.
\newblock


\bibitem[Jim{\'e}nez and Lin(2001)]%
        {jimenez01}
\bibfield{author}{\bibinfo{person}{Daniel~A. Jim{\'e}nez} {and} \bibinfo{person}{Calvin Lin}.} \bibinfo{year}{2001}\natexlab{}.
\newblock \showarticletitle{Dynamic Branch Prediction with Perceptrons}. In \bibinfo{booktitle}{\emph{HPCA}}. \bibinfo{pages}{197--206}.
\newblock


\bibitem[Knuth(1974)]%
        {knuth74}
\bibfield{author}{\bibinfo{person}{Donald~E. Knuth}.} \bibinfo{year}{1974}\natexlab{}.
\newblock \showarticletitle{Structured Programming with go to Statements}.
\newblock \bibinfo{journal}{\emph{Comput. Surveys}} (\bibinfo{year}{1974}).
\newblock


\bibitem[Kung and Robinson(1981)]%
        {kung81}
\bibfield{author}{\bibinfo{person}{H.~T. Kung} {and} \bibinfo{person}{John~T. Robinson}.} \bibinfo{year}{1981}\natexlab{}.
\newblock \showarticletitle{On Optimistic Methods for Concurrency Control}.
\newblock \bibinfo{journal}{\emph{ACM Transactions on Database Systems}} \bibinfo{volume}{6}, \bibinfo{number}{2} (\bibinfo{year}{1981}), \bibinfo{pages}{213--226}.
\newblock
\href{https://doi.org/10.1145/319566.319567}{doi:\nolinkurl{10.1145/319566.319567}}


\bibitem[Lamport(1998a)]%
        {lamport}
\bibfield{author}{\bibinfo{person}{Leslie Lamport}.} \bibinfo{year}{1998}\natexlab{a}.
\newblock \showarticletitle{The Part-Time Parliament}.
\newblock \bibinfo{journal}{\emph{ACM Transactions on Computer Systems}} (\bibinfo{year}{1998}).
\newblock


\bibitem[Lamport(1998b)]%
        {lamport98}
\bibfield{author}{\bibinfo{person}{Leslie Lamport}.} \bibinfo{year}{1998}\natexlab{b}.
\newblock \showarticletitle{The Part-Time Parliament}.
\newblock \bibinfo{journal}{\emph{ACM TOCS}} (\bibinfo{year}{1998}).
\newblock


\bibitem[Lamport(2002)]%
        {Lamport02}
\bibfield{author}{\bibinfo{person}{Leslie Lamport}.} \bibinfo{year}{2002}\natexlab{}.
\newblock \bibinfo{booktitle}{\emph{Specifying Systems: The TLA+ Language and Tools for Hardware and Software Engineers}}.
\newblock \bibinfo{publisher}{Addison-Wesley}.
\newblock


\bibitem[Lampson(1974)]%
        {lampson71}
\bibfield{author}{\bibinfo{person}{Butler~W. Lampson}.} \bibinfo{year}{1974}\natexlab{}.
\newblock \showarticletitle{Protection}.
\newblock \bibinfo{journal}{\emph{ACM Operating Systems Review}} (\bibinfo{year}{1974}).
\newblock


\bibitem[Lampson(1983)]%
        {lampson83}
\bibfield{author}{\bibinfo{person}{Butler~W. Lampson}.} \bibinfo{year}{1983}\natexlab{}.
\newblock \showarticletitle{Hints for Computer System Design}.
\newblock \bibinfo{journal}{\emph{ACM Operating Systems Review}} (\bibinfo{year}{1983}).
\newblock


\bibitem[Lattner and Adve(2004)]%
        {lattner04}
\bibfield{author}{\bibinfo{person}{Chris Lattner} {and} \bibinfo{person}{Vikram Adve}.} \bibinfo{year}{2004}\natexlab{}.
\newblock \showarticletitle{LLVM: A Compilation Framework for Lifelong Program Analysis \& Transformation}. In \bibinfo{booktitle}{\emph{CGO}}.
\newblock


\bibitem[Lawson et~al\mbox{.}(1979)]%
        {lawson79blas}
\bibfield{author}{\bibinfo{person}{C.~L. Lawson}, \bibinfo{person}{R.~J. Hanson}, \bibinfo{person}{D.~R. Kincaid}, {and} \bibinfo{person}{F.~T. Krogh}.} \bibinfo{year}{1979}\natexlab{}.
\newblock \showarticletitle{Basic Linear Algebra Subprograms for Fortran Usage}.
\newblock \bibinfo{journal}{\emph{ACM Trans. Math. Software}} (\bibinfo{year}{1979}).
\newblock


\bibitem[Levin et~al\mbox{.}(1975)]%
        {levin75}
\bibfield{author}{\bibinfo{person}{R. Levin}, \bibinfo{person}{E. Cohen}, \bibinfo{person}{W. Corwin}, \bibinfo{person}{F. Pollack}, {and} \bibinfo{person}{W. Wulf}.} \bibinfo{year}{1975}\natexlab{}.
\newblock \showarticletitle{Policy/Mechanism Separation in Hydra}. In \bibinfo{booktitle}{\emph{SOSP}}.
\newblock


\bibitem[Lieberman(2001)]%
        {lieberman01}
\bibfield{author}{\bibinfo{person}{Henry Lieberman}.} \bibinfo{year}{2001}\natexlab{}.
\newblock \bibinfo{booktitle}{\emph{Your Wish is My Command: Programming by Example}}.
\newblock \bibinfo{publisher}{Morgan Kaufmann}.
\newblock


\bibitem[Matsakis and Klock(2014)]%
        {matsakis14}
\bibfield{author}{\bibinfo{person}{Nicholas~D. Matsakis} {and} \bibinfo{person}{Felix Klock}.} \bibinfo{year}{2014}\natexlab{}.
\newblock \showarticletitle{The Rust Language}. In \bibinfo{booktitle}{\emph{ACM SIGAda}}.
\newblock


\bibitem[Morris(1989)]%
        {morris89}
\bibfield{author}{\bibinfo{person}{Robert~T. Morris}.} \bibinfo{year}{1989}\natexlab{}.
\newblock \showarticletitle{A Tour of the Worm}.
\newblock \bibinfo{journal}{\emph{USENIX}} (\bibinfo{year}{1989}).
\newblock


\bibitem[Parnas(1979)]%
        {Parnas79}
\bibfield{author}{\bibinfo{person}{David~L. Parnas}.} \bibinfo{year}{1979}\natexlab{}.
\newblock \showarticletitle{Designing Software for Ease of Extension and Contraction}.
\newblock \bibinfo{journal}{\emph{IEEE Transactions on Software Engineering}} (\bibinfo{year}{1979}).
\newblock


\bibitem[Paxson(1999)]%
        {Paxson99}
\bibfield{author}{\bibinfo{person}{Vern Paxson}.} \bibinfo{year}{1999}\natexlab{}.
\newblock \showarticletitle{End-to-End Internet Packet Dynamics}.
\newblock \bibinfo{journal}{\emph{IEEE/ACM Transactions on Networking}} (\bibinfo{year}{1999}).
\newblock


\bibitem[Petersen et~al\mbox{.}(1997)]%
        {petersen97}
\bibfield{author}{\bibinfo{person}{K. Petersen}, \bibinfo{person}{M. Spreitzer}, \bibinfo{person}{D. Terry}, \bibinfo{person}{M. Theimer}, {and} \bibinfo{person}{A. Demers}.} \bibinfo{year}{1997}\natexlab{}.
\newblock \showarticletitle{Flexible Update Propagation for Weakly Consistent Replication}. In \bibinfo{booktitle}{\emph{SOSP}}.
\newblock


\bibitem[Pirahesh et~al\mbox{.}(1992)]%
        {pirahesh92}
\bibfield{author}{\bibinfo{person}{Hamid Pirahesh}, \bibinfo{person}{Joseph~M. Hellerstein}, {and} \bibinfo{person}{Waqar Hasan}.} \bibinfo{year}{1992}\natexlab{}.
\newblock \showarticletitle{Extensible/Rule Based Query Rewrite Optimization in Starburst}. In \bibinfo{booktitle}{\emph{SIGMOD}}.
\newblock


\bibitem[Popek and Goldberg(1974)]%
        {popek74}
\bibfield{author}{\bibinfo{person}{Gerald~J. Popek} {and} \bibinfo{person}{Robert~P. Goldberg}.} \bibinfo{year}{1974}\natexlab{}.
\newblock \showarticletitle{Formal Requirements for Virtualizable Third Generation Architectures}.
\newblock \bibinfo{journal}{\emph{Commun. ACM}} (\bibinfo{year}{1974}).
\newblock


\bibitem[Ritchie and Thompson(1974)]%
        {ritchie74}
\bibfield{author}{\bibinfo{person}{Dennis~M. Ritchie} {and} \bibinfo{person}{Ken Thompson}.} \bibinfo{year}{1974}\natexlab{}.
\newblock \showarticletitle{The {UNIX} Time-Sharing System}.
\newblock \bibinfo{journal}{\emph{Commun. ACM}} (\bibinfo{year}{1974}).
\newblock


\bibitem[Saltzer et~al\mbox{.}(1984)]%
        {saltzer84}
\bibfield{author}{\bibinfo{person}{J.~H. Saltzer}, \bibinfo{person}{D.~P. Reed}, {and} \bibinfo{person}{D.~D. Clark}.} \bibinfo{year}{1984}\natexlab{}.
\newblock \showarticletitle{End-to-End Arguments in System Design}.
\newblock \bibinfo{journal}{\emph{ACM Transactions on Computer Systems}} (\bibinfo{year}{1984}).
\newblock


\bibitem[Saltzer and Schroeder(1975)]%
        {saltzer75}
\bibfield{author}{\bibinfo{person}{Jerome~H. Saltzer} {and} \bibinfo{person}{Michael~D. Schroeder}.} \bibinfo{year}{1975}\natexlab{}.
\newblock \showarticletitle{The Protection of Information in Computer Systems}.
\newblock \bibinfo{journal}{\emph{Proc. IEEE}} (\bibinfo{year}{1975}).
\newblock


\bibitem[Selinger et~al\mbox{.}(1979)]%
        {selinger79}
\bibfield{author}{\bibinfo{person}{Patricia~G. Selinger}, \bibinfo{person}{Morton~M. Astrahan}, \bibinfo{person}{Donald~D. Chamberlin}, \bibinfo{person}{Raymond~A. Lorie}, {and} \bibinfo{person}{Thomas~G. Price}.} \bibinfo{year}{1979}\natexlab{}.
\newblock \showarticletitle{Access Path Selection in a Relational Database Management System}. In \bibinfo{booktitle}{\emph{SIGMOD}}.
\newblock


\bibitem[Stepanov and Lee(1994)]%
        {stepanov94}
\bibfield{author}{\bibinfo{person}{Alexander~A. Stepanov} {and} \bibinfo{person}{Meng Lee}.} \bibinfo{year}{1994}\natexlab{}.
\newblock \bibinfo{booktitle}{\emph{The Standard Template Library}}.
\newblock \bibinfo{type}{{T}echnical {R}eport}. \bibinfo{institution}{Hewlett-Packard Laboratories}.
\newblock
\urldef\tempurl%
\url{https://stepanovpapers.com/STL/DOC.PDF}
\showURL{%
\tempurl}


\bibitem[Stonebraker and Rowe(1986)]%
        {stonebraker86}
\bibfield{author}{\bibinfo{person}{Michael Stonebraker} {and} \bibinfo{person}{Lawrence~A. Rowe}.} \bibinfo{year}{1986}\natexlab{}.
\newblock \showarticletitle{The Design of {POSTGRES}}. In \bibinfo{booktitle}{\emph{SIGMOD}}.
\newblock


\bibitem[Whaley and Dongarra(1998)]%
        {whaley98}
\bibfield{author}{\bibinfo{person}{R.~Clint Whaley} {and} \bibinfo{person}{Jack~J. Dongarra}.} \bibinfo{year}{1998}\natexlab{}.
\newblock \showarticletitle{Automatically Tuned Linear Algebra Software}. In \bibinfo{booktitle}{\emph{SC}}.
\newblock


\bibitem[Zimmermann(1980)]%
        {zimmermann80}
\bibfield{author}{\bibinfo{person}{Hubert Zimmermann}.} \bibinfo{year}{1980}\natexlab{}.
\newblock \showarticletitle{{OSI} Reference Model -- The {ISO} Model of Architecture for Open Systems Interconnection}.
\newblock \bibinfo{journal}{\emph{IEEE Transactions on Communications}} (\bibinfo{year}{1980}).
\newblock


\end{thebibliography}

\end{document}